\documentclass[11pt, letter]{article}
\usepackage{jheppub}
\usepackage{avant} 
\usepackage[dvipsnames]{xcolor}
\usepackage{amsmath,epsf,amssymb,mathtools,latexsym,amsthm,setspace,array,pifont,hyperref,amsfonts,dsfont,cancel,braket,parskip,slashed}
\usepackage[none]{hyphenat} 
\usepackage{graphicx,verbatim}
\usepackage{float}
\usepackage{tikz} 
\usepackage{tikz-feynman, contour}
\tikzfeynmanset{compat=1.1.0}
\usetikzlibrary{shapes,arrows,positioning,automata,backgrounds,calc,er,patterns}
\makeatletter
\usepackage{todonotes}
\usepackage{enumitem,amssymb}
\newlist{todolist}{itemize}{2}
\setlist[todolist]{label=$\square$}
\usepackage{pifont}

\graphicspath{{Pictures/}} 
\newcommand{\nn}{\nonumber}

\newcommand{\pd}{\partial}

\newcommand{\A}{\alpha}
\newcommand{\B}{\beta}

\newcommand{\cl}[1]{\mathcal{#1}}

\def\prd{\ref@{Phys.~Rev.~D}}        
\newcommand{\Tr}[1]{\text{Tr}\left(#1\right)}
\usepackage{tcolorbox}
\definecolor{airforceblue}{rgb}{0.36, 0.54, 0.66}
\definecolor{azure}{rgb}{0.0, 0.5, 1.0}
\newtcolorbox{tdbox}{colback=airforceblue!40!white,colframe=azure!90!black} 
\newcommand{\td}[1]{
	\if\notesOn1
	\begin{tdbox}
		#1
	\end{tdbox}
	\fi
}
\newcommand{\com}[1]{
	\if\commentsOn1
	\begin{tdbox}
		#1
	\end{tdbox}
	\fi
}

\hypersetup{
	colorlinks = true,
	linkcolor = Mahogany,
	anchorcolor = Mahogany,
	citecolor = Mahogany,
	filecolor = Mahogany,
	urlcolor = Mahogany
}

\def\notesOn{0}
\def\commentsOn{1}
\tikzset{
	graviton/.style={
		double,
		decoration={snake, aspect=0.75, mirror, segment length=1.5mm},
		decorate
	}
}

\tikzfeynmanset{
	bigblob/.style={
		shape=circle,
		draw=blue,
		fill=red}
}

\title{Scattering Amplitudes and the Double Copy in Topologically Massive Theories}
\author{Nathan Moynihan}
\affiliation{High Energy Physics, Cosmology \& Astrophysics Theory (HEPCAT) Group\\
	and	The Laboratory for Quantum Gravity \& Strings (QGASLab),\\
	Department of Mathematics and Applied Mathematics, University of Cape Town\\
	Rondebosch, Cape Town 7700, South Africa}
\emailAdd{nathantmoynihan@gmail.com}
\abstract{
	Using the principles of the modern scattering amplitudes programme, we develop a formalism for constructing the amplitudes of three-dimensional topologically massive gauge theories and gravity. Inspired by recent developments in four dimensions, we construct the three-dimensional equivalent of $x$-variables, first defined in \cite{Arkani-Hamed:2017jhn}, for conserved matter currents coupled to topologically massive gauge bosons or gravitons. Using these, we bootstrap various matter-coupled gauge-theory and gravitational scattering amplitudes, and conjecture that topologically massive gauge theory and topologically massive gravity are related by the double copy. To motivate this idea further, we show explicitly that the Landau gauge propagator on the gauge theory side double copies to the de Donder gauge propagator on the gravity side.
}

\begin{document}
\maketitle
\newpage 
\section{Introduction}
Three dimensional quantum field theories are of interest in many areas of physics, from condensed matter to string theory and the AdS/CFT correspondence. In some sense, quantum field theories in three dimensions have a much richer structure than their four dimensional cousins. This is partly due to the fact that Chern-Simons terms are permitted in odd dimensions, which in $2+1$ dimensions allows for the presence of particles with fractional statistics (anyons). One particularly interesting class of three dimensional QFT's are \textit{topologically massive} theories \cite{Deser:1981wh}, which include gauge theories and gravity, both with a topological mass related to the Chern-Simons level number. Topologically massive theories are parity-odd and have the remarkable property of being massive \textit{and} gauge invariant, and could be useful as a toy model for understanding gauge theories and gravity in $3+1$ dimensions, giving rise to e.g. the BTZ black hole \cite{Banados:1992wn} as a playground to test AdS/CFT. 

In this paper, we will explore this class of theories by bootstrapping their scattering using the modern on-shell approach. Scattering amplitudes in three dimensions are well studied using spinor-helicity techniques \cite{Chiou:2005jn,Agarwal:2008pu,Gang:2010gy,Lipstein:2012kd,Bargheer:2012cp,Brandhuber:2012un,Huang:2010rn,Bargheer:2012gv,Agarwal:2011hb,Agarwal:2012jj,Lee:2010du,Agarwal:2013tpa,Huang:2012wr,Chin:2015qza,Adamo:2017xaf}, and we will utilise this technology extensively. Inspired by recent progress in four dimensions, we will develop a formalism in three-dimensions for constructing the amplitudes directly, in particular by constructing the three-dimensional equivalent of $x$-variables, originally defined in four dimensions in \cite{Arkani-Hamed:2017jhn}. In particular we will find that this is extremely useful in the context of scattering in topologically massive gravity \cite{Dengiz:2013hka}, whose complicated Feynman rules make calculating scattering amplitudes horribly cumbersome. In four dimensions, ratios of $x$-variables encode both `electric' and `magnetic' degrees of freedom, where the magnetic degrees of freedom are encoded by a Levi-Civita term coupled to a Dirac string \cite{Caron-Huot:2018ape,Huang:2019cja,Moynihan:2020gxj}. In three-dimensions, we find that the magnetic degrees of freedom are encoded in the $x$-variables in none other than the Chern-Simons term. Crucial to this construction is the fact that the little group of massless fields in $D$ dimensions is the same as for \textit{massive} fields in $D-1$ dimensions, allowing us to use much of the technology built for four dimensional massless amplitudes to construct three dimensional massive ones.

Along the way we will motivate the idea that topologically massive gauge and gravity theories are related via the \textit{double copy}, a correspondence between gauge and gravitational theories originally formulated in the context of massless scattering amplitudes in four dimensions \cite{Bern:2008qj,Bern:2010yg,Bern:2010ue}. At this point, the double copy has been extended to include massive particles \cite{Moynihan:2017tva,Johansson:2019dnu,Momeni:2020vvr,Johnson:2020pny,Adamo:2020qru,Haddad:2020tvs} and purely classical solutions \cite{Monteiro:2014cda,Luna:2015paa,Luna:2016due,Goldberger:2016iau,Goldberger:2017frp,Goldberger:2017vcg,Goldberger:2017ogt,Luna:2016hge,Luna:2017dtq,Shen:2018ebu,Levi:2018nxp,Plefka:2018dpa,Cheung:2018wkq,Carrillo-Gonzalez:2018pjk,Monteiro:2018xev,Plefka:2019hmz,Maybee:2019jus,PV:2019uuv,Carrillo-Gonzalez:2019aao,Bautista:2019evw,Moynihan:2019bor,Bah:2019sda,Goldberger:2019xef,Kim:2019jwm,Banerjee:2019saj,Alawadhi:2019urr,Alfonsi:2020lub,Bahjat-Abbas:2020cyb,Luna:2020adi,Cristofoli:2020hnk} including in three spacetime dimensions \cite{Carrillo-Gonzalez:2017iyj,CarrilloGonzalez:2019gof,Gumus:2020hbb}. In three dimensions, however, the double copy is not so straightforward, since we almost immediately hit a road block in the sense that massless Yang-Mills theories only propagate scalar degrees of freedom and 3D general relativity propagates none, being purely topological (the Weyl tensor vanishes). That being said, strictly speaking the double copy relates Yang-Mills theories to dilaton-axion gravity rather than simply general relativity, and it has been shown that the \textit{classical} double copy does indeed hold in three dimensions \cite{Carrillo-Gonzalez:2017iyj,CarrilloGonzalez:2019gof,Gumus:2020hbb}, relating the gauge theory solution to a dilaton-like solution. Furthermore, there are many examples of supergravity theories that are double copies, and this is equally true in three dimensions where many different massless super-Yang Mills amplitudes double copy to their supergravity counterparts \cite{Bargheer:2012gv,Huang:2012wr}.  

An interesting observation that one can make about gravity in general dimensions is that massless degrees of freedom in $D$ dimensions are in some sense mapped to massive degrees of freedom in $D-1$ dimensions. For example, the massless graviton propagator in $D$ dimensions is equal to the massive graviton propagator in $D-1$ dimensions \cite{Hinterbichler:2011tt}: massive particles in $D-1$ dimensions have the same number of degrees of freedom as massless particles in $D$ dimensions\footnote{To see this, we note that massless gravitons have $\frac12D(D-3)$ degrees of freedom while massive gravitons have $\frac12(D+1)(D-2)$. We see then that massless gravitons in four dimensions and massive gravitons in three both have 2 degrees of freedom.}. As alluded to earlier, for both four dimensional massless theories and three dimensional massive theories the little group is $U(1)$. Together with dimensional analysis and locality, this is all we need to construct a large class of massive particle scattering amplitudes in three dimensions.

We achieve this by noting that spinor-helicity variables in $2+1$ dimensions naturally encode the spin degrees of freedom of topologically massive particles, just as their four dimensional counterparts encode helicities. As alluded to earlier, this allows us to classify amplitudes according to their little group structure --- $U(1)$ for massive particles --- and bootstrap the amplitudes directly by demanding that little group covariance, locality and dimensional consistency are respected. We find that the double copy is manifest at the two and three particle level, in the off-shell propagator and in a selection of on-shell amplitudes. We confirm a number of these results by computing the amplitudes directly from the Feynman rules in appendix \ref{feynappendix}.

\section{Scattering Amplitudes in 2+1 Dimensions}
In order to construct amplitudes in $2+1$ dimensions, we will make extensive use of the spinor-helicity formalism, which in $D=3$ consists only of angle brackets $\ket{i}$ and spinor contractions of the form $\braket{ij} = \lambda_i^\A\lambda_j^\B\epsilon_{\B\A}$. In $D$ spacetime dimensions, the Lorentz group is given by $SO(D-1,1)$.  In order to construct amplitudes in $D = 3$, we need to know about the little group structure that corresponds to either massive or massless particles. For massive particles in $D$ dimensions, the little group is $SO(D-1)$, and for massless it's $E(D-2)$. For $D=3$ then, this gives $SO(2) \sim U(1)$ as the little group for massive particles and $E(1) \sim \mathbb{R}$ for massless. This is easily seen by considering a massive particle in its rest frame or a massless particle boosted along a given axis.

In order to build scattering amplitudes, we need to consider the possible one-particle states that can exist in three dimensions, presumably having spin-$s$ and mass $m$ (along with whatever other quantum numbers may be present). As usual, these will be classified by the Casimirs of the Poincar\'e group $SO(2,1)$, $p^2$ and $S^2$, which are related to the invariant mass and the spin, respectively. The spin in this case characterised by the Pauli-Lubanski pseudoscalar
\begin{equation}
\sigma = \epsilon^{\mu\nu\rho}\sigma_{\mu\nu}p_\rho = J\cdot p
\end{equation}
where $\sigma_{\mu\nu} = \frac{i}{4}[\gamma_\mu,\gamma_\nu] = \frac12\epsilon_{\mu\nu\rho}\gamma^\rho$ is the angular momentum tensor and $p_\rho$ the three-momentum. 

Massless particles in $D = 3$ only come in two types: scalars and (spinless) fermions \cite{Binegar:1981gv,Jackiw:1990ka}. This means that familiar concepts in $D = 4$ gauge theories, mediated by massless spin-1 particles, do not necessarily exist for massless theories in $D= 3$. This is even worse for gravity, where massless gravitons are purely topological and have no degrees of freedom. The situation is far less dire for generic massive states, which have two degrees of freedom on-shell and possible spins $s = \pm\frac12,\pm1,\pm\frac32,\pm2,\cdots$, and are essentially classified in the same way as massless particles in $D=4$ (i.e. by their helicity). In $D = 4$, the spinor helicity formalism encapsulates the helicity degrees of freedom for massless particles, and we can do the same in $D = 3$ for massive states of spin $\pm s$.

In this paper, we are interested in topologically massive gauge theories, where the sign of the spin is given by the sign of the Chern-Simons level number $k$, which in turn is related to the topological mass as $sgn({s}) = \frac{k}{|k|} = \frac{m}{|m|}$ \cite{Deser:1981wh}.

Momentum vectors in three dimensions have a spinor decomposition given by
\begin{equation}
p^\mu\sigma_\mu^{\A\B} = p^{\A\B} = \lambda^{(\A}\bar{\lambda}^{\B)},
\end{equation}
where we have followed the conventions set out in appendix \ref{conventions}, and the spinors are given explicitly as
\begin{equation}
\lambda_\A = \frac{1}{\sqrt{p_0-p_1}}\begin{pmatrix}
p_2 -im\\
p_1 - p_0 
\end{pmatrix},~~~~~\bar{\lambda}_\A = -\frac{1}{\sqrt{p_0-p_1}}\begin{pmatrix}
p_2 +im\\
p_1 - p_0 
\end{pmatrix}.
\end{equation}
We will often use angle brackets as a shorthand for the spinor inner product, i.e.
\begin{equation}
	\epsilon^{\alpha\beta}\lambda_{i\beta}\lambda_{j\alpha} \equiv \braket{ij},~~~~~\epsilon^{\alpha\beta}\lambda_{i\beta}\bar{\lambda}_{j\alpha} \equiv \braket{i\bar{j}}.
\end{equation}
We observe then that these particular spinor helicity variables are very natural objects to consider for topologically massive states in three dimensions. This is because each spinor $\lambda$ or $\bar{\lambda}$ differs only by the sign of the imaginary part $im = isgn(s)|m|$, relative to the sign of the momentum components. This means that $\lambda$ ($\bar{\lambda}$) encodes the negative (positive) spin-degrees of freedom for topologically massive theories, in analogy to how $\lambda$ and $\tilde{\lambda}$ encodes the helicity states for massless theories in 4D.

We can use this information to infer how amplitudes for a given spin should behave under little group transformations, again analogous to 4D, i.e. that
\begin{equation}
\cl{A}_n(e^{i\omega}\lambda_i, e^{-i\omega}\bar{\lambda}_i) \rightarrow e^{2is_i\omega}\cl{A}(\lambda_i, \bar{\lambda}_i).
\end{equation}

Kinematics in three dimensions are more constrained than in four dimensions. We can express dot products of momentum vectors in terms of spinors as
\begin{equation}
2p_i\cdot p_j = \frac12\left(\braket{ij}\braket{\bar{i}\bar{j}} + \braket{i\bar{j}}\braket{\bar{i}j}\right),
\end{equation}
which immediately tells us that $p_i^2 = \frac14\braket{i\bar{i}}^2 = -m_i^2$. Furthermore, we see that products of masses are now subject to the Schouten identity, and we find that they can be written in terms of spinors as
\begin{equation}
m_im_j = -\frac14(\braket{i\bar{i}}\braket{j\bar{j}}) = -\frac14({\braket{ij}\braket{\bar{i}\bar{j}} - \braket{i\bar{j}}\braket{\bar{i}j}})
\end{equation}
The Mandelstam variables are given by
\begin{equation}
s_{ij} = -(p_i + p_j)^2 = m_i^2+m_j^2 -2p_i\cdot p_j.
\end{equation}
We also note that we can again use the Schouten identity to write
\begin{equation}
2im_i\braket{jk} = \braket{i\bar{i}}\braket{jk} = \braket{j\bar{i}}\braket{ik}-\braket{ji}\braket{\bar{i}k}.
\end{equation}
Momentum conservation for massive 3D momentum is given by
\begin{equation}
\sum_{i = 1}^n \left(\ket{\bar{i}}\bra{i} + \ket{i}\bra{\bar{i}}\right) = 0.
\end{equation}

\subsection{Three Particle Amplitudes}
There are no massless three-particle amplitudes in three spacetime dimensions, since there are no Lorentz invariant objects that one can write down that aren't zero\footnote{In principle, we could have a constant three-particle amplitude here, corresponding to a term $\sim g\phi^3$, where $[g] = 3/2$, however we won't consider such a theory here.}, since $s_{ij} = \braket{ij}^2 = 0$ for all $i,j$. Massive three-particle amplitudes are allowed however, and for self-interacting theories we can construct them directly from symmetry considerations. There are two possible spin configurations we can write down at the three particle level: all same-spin $(s_1 = s_2 = s_3)$ or one different (e.g. $s_1 = s_2 = -s_3$), with all other amplitudes obtainable by complex conjugation or particle relabelling. 

Scattering amplitudes in spacetime dimension $D$ with $n$ external legs have mass dimension $[\cl{A}_n] = \frac{n}{2}(2-D) + D$. In $D=3$ then, this implies a fractional mass dimension for amplitudes with an odd numbers of legs and specifically that 3-particle amplitudes must have mass dimension $3/2$. In addition to this, the three-particle amplitudes must have the correct little group scaling and they must vanish in the all massless limit, although this last constraint is trivial. In order to expose the little group covariance of a given amplitude, we can act on it with the spin-operator
\begin{equation}\label{spinop}
S_i = \lambda_i\frac{\pd}{\pd\lambda_i} - \bar{\lambda}_i\frac{\pd}{\pd\bar{\lambda}_i},
\end{equation}
which returns the spin for a given particle $i$. The simplest object that satisfies $S_i\cl{A}_3[i^{\pm s}] = \pm s\cl{A}_3$ and can be built from only $\lambda, \bar{\lambda}$ and $m$ is given by
\begin{equation}
\cl{A}_3[1_a^{s},2_b^{s},\bar{3}_c^{s}] = g_{abc}\left(i\frac{\braket{12}\braket{2\bar{3}}\braket{\bar{3}1}}{m^2}\right)^s
\end{equation}
This can be related to more complicated amplitudes (more familiar from dimensional reduction or via the Feynman rules) via judicious use of the Schouten identity to find 
\begin{align}
i\frac{\braket{12}\braket{2\bar{3}}\braket{\bar{3}1}}{m^2} &= -\frac12\left(\frac{\braket{12}^2\braket{\bar{3}|p_1|\bar{3}} + \braket{1\bar{3}}^2\braket{2|p_1|2}}{m^3}\right),
\end{align}
where other spin configurations can be found by barring and unbarring the different spinors and the coupling is required to have mass dimension $[g_{abc}] = 3/2 - s$ for consistency. 

We see immediately that for $s=1$ the coupling must inherit the anti-symmetry properties of the spinors to respect Bose symmetry, i.e. that $g_{abc} = -g_{acb}$, meaning this self-interaction must come from a non-Abelian theory. For $s=2$ there is no such requirement, however do note that the coupling must now have dimension $[g_{abc}] = -1/2$ and as such should describe a gravitational theory. While it is certainly interesting to study both pure gauge theory and pure gravity, in this paper we will restrict ourselves to considering these fields when coupled to matter. Specifically, we will consider matter particles minimally coupled to either topologically massive gauge fields or topologically massive gravitons and define the three-dimensional equivalent of the $x$-variables first defined in \cite{Arkani-Hamed:2017jhn}. We will then show that gauge and gravity mediated interactions are related by simply squaring the $x$-variables, as is the case in four dimensions. 

In the four dimensional case, the $x$-variables are defined as proportionality constants between massless basis spinors when the mass of each external particle are equal. While in this case we have no massless basis spinors, we can still consider same-mass matter particles, which essentially just means we are considering spin-$s$ particles coupled to conserved matter currents. Considering the amplitude above, we can easily couple scalars to the topologically massive spin $s$ boson by taking the `scalar' limit of two of the particles (by barring half of the particles spinors). For example, the scalar-scalar-massive gauge boson amplitude is given by
\begin{equation}
\cl{A}_3[1^{0},2^{0},3^{s}] = g\frac{\braket{\bar{3}|p_1|\bar{3}}^s}{m^s} = g(mx_{12})^{s},~~~~~\cl{A}_3[1^{0},2^{0},3^{-s}] = g\frac{\braket{3|p_1|3}^s}{m^s} = g\frac{m^s}{x_{12}^s}
\end{equation}

Pleasingly, this is also what we would obtain by dimensional reduction of the four-dimensional $x$-variables, and so we define  $x$ by analogy as
\begin{equation}
x_{12} = \frac{\braket{\bar{3}|u_1|\bar{3}}}{2\sqrt{-p_3^2}},~~~~~\frac{1}{x_{12}} = \frac{\braket{3|u_1|3}}{2\sqrt{-p_3^2}},~~~~~u_i = \frac{p_i}{m_i}.
\end{equation}
To see that these are inverse to one another, we note that
\begin{equation}
\braket{3|u_1|3}\braket{\bar{3}|u_1|\bar{3}} = \braket{3|u_1p_3u_1|\bar{3}} + im_3\braket{3|u_1u_1|\bar{3}} = 4m_3^2, 
\end{equation}
where we have used the identity in eq. \eqref{fourident}.

We can then write a generalisation for a topologically massive gauge field of spin $s$ coupled to a spin $s'$ matter current\footnote{We note that this formula generalises the one expected for a massless gauge boson and recovers it in the appropriate limit, e.g.
\begin{equation}
\lim_{m_s\rightarrow 0}\cl{A}_3[1^{s'},2^{s'},3^{s}] = g(m_1x_{12})^{s}\frac{\braket{12}^{2s'}}{m_1^{2s'}}.
\end{equation}  }, noting that these should satisfy $\lim_{m_s,m_{s'}\rightarrow 0}\cl{A}_3 = 0$,
\begin{align}
\cl{A}_3[1^{s'},2^{s'},3^{s}] &= g(m_1x_{12})^{s}\frac{2^{s'}\braket{12}^{2s'}}{(m_1^2 + p_1\cdot p_2)^{s'}},\\\cl{A}_3[1^{s'},2^{s'},3^{-s}] &= g\left(\frac{m_1}{x_{12}}\right)^{s}\frac{2^{s'}\braket{12}^{2s'}}{(m_1^2 + p_1\cdot p_2)^{s'}},
\end{align}
where the different spin configurations can be chosen by barring the appropriate spinor, provided the absolute value of the spin remains the same for both matter particles.

One important thing to note about topologically massive theories is that they are not parity symmetric. In order to consider a parity symmetric theory, we could follow Ref. \cite{Deser:1981wh} and consider a \textit{doublet} of particles that are identical in every way except for their mass, which differs by a sign as $m = \pm|m|$. This would amount to summing over the distinct propagating particles (which would now carry both spins $\pm s$), which would eliminate parity-specific (i.e. Levi-Civita) terms, effectively giving pure Yang-Mills contributions. We will \textit{not} consider such parity symmetric theories here since we wish to consider the full topologically massive theory, which essentially means we will not sum over any $x$-ratios that we encounter. However, it is interesting to consider how deforming the parity symmetric theory (e.g. 3D Yang-Mills) might lead to a parity violating one by, for example, deforming the $x$-ratios. This has been especially fruitful in four dimensions \cite{Guevara:2017csg,Guevara:2018wpp,Guevara:2019fsj,Huang:2019cja,Moynihan:2020gxj}, and will be explored in detail elsewhere \cite{Emond:2021}.

\section{Four Particle Amplitudes}
\subsection{Scalar Scattering}
The simplest four particle amplitude we can consider computing is two scalars exchanging a gauge boson with coupling $g = \sqrt{2}e$. This is given by
\begin{align}
\vcenter{\hbox{
		\begin{tikzpicture}[scale=0.5]
		]
		\begin{feynman}  
		\vertex (a) at (-4,2) {$2$};
		\vertex (b) at (-4,-2) {$1$};
		\vertex (c) at (2,-2) {$4$};
		\vertex (d) at (2,2) {$3$};
		\vertex (r) at (0,0);
		\vertex (l) at (-2,0) ;
		\diagram* {
			(a) -- [plain] (l) -- [photon, momentum={$q$}] (r) -- [plain] (d),
			(b) -- [plain] (l) -- [photon] (r) -- [plain] (c),
		};
		\end{feynman}
		\end{tikzpicture}}} &= \frac{2e^2m^2_\phi}{(s-m_\gamma^2)}\left(\frac{x_{12}}{x_{34}}\right)\\
&= e^2\frac{\braket{\bar{q}|p_1|\bar{q}}\braket{q|p_4|q}}{2s(s-m_\gamma^2)}\\
&= e^2\frac{2m^2(p_1-p_2)\cdot p_4 + 8im_\gamma\epsilon(p_1,p_2,p_3)}{2s(s-m_\gamma^2)}\\
&= e^2\frac{s(t-u) + 4im\epsilon(p_1,p_2,p_3)}{s(s-m_\gamma^2)}
\end{align}
where we have used eq. \eqref{fourident} and made crossing symmetry manifest.
Adding the other channels then gives the full amplitude
\begin{align}
\cl{A}_4[1,2,3,4] = & e^2\frac{s(t-u) + 4im_\gamma\epsilon(p_1,p_2,p_3)}{s(s-m_\gamma^2)} + e^2\frac{u(t-s) + 4im_\gamma\epsilon(p_1,p_2,p_3)}{u(u-m_\gamma^2)}\nn\\
&+ e^2\frac{t(u-s) - 4im_\gamma\epsilon(p_1,p_2,p_3)}{t(t-m_\gamma^2)}
\end{align}

The topologically massive gravity version, with $g = \kappa/2$, is given by squaring the $x$-ratio
\begin{align}
\vcenter{\hbox{
		\begin{tikzpicture}[scale=0.5]
		]
		\begin{feynman}  
		\vertex (a) at (-4,2) {$2$};
		\vertex (b) at (-4,-2) {$1$};
		\vertex (c) at (2,-2) {$4$};
		\vertex (d) at (2,2) {$3$};
		\vertex (r) at (0,0);
		\vertex (l) at (-2,0) ;
		\diagram* {
				(a) -- [plain] (l) -- [graviton] (r) -- [plain] (d),
				(b) -- [plain] (l) -- [graviton] (r) -- [plain] (c),
		};
		\end{feynman}
		\end{tikzpicture}}} &= \frac{\kappa^2m^4_\phi}{4(s-m_\gamma^2)}\left(\frac{x_{12}}{x_{34}}\right)^2\\
&= \kappa^2\frac{\braket{\bar{q}|p_1|\bar{q}}^2\braket{q|p_4|q}^2}{16s^2(s-m_\gamma^2)}\\
&= \frac{\kappa^2}{8}\frac{s^2(t-u)^2 - 16s\Gamma(p_1,p_2,p_3) + 8ims(t-u)\epsilon(p_1,p_2,p_3)}{s^2(s-m_\gamma^2)}\\
&= \frac{\kappa^2}{8}\frac{s^2(t^2+u^2-6tu) + 8ims(t-u)\epsilon(p_1,p_2,p_3)}{s^2(s-m_\gamma^2)},
\end{align}
where $\Gamma(p_1,p_2,p_3)$ is the Gram determinant.  We find then that the full amplitude is
\begin{align}
\cl{M}_4[1,2,3,4] = &-\frac{\kappa^2}{8}\left(\frac{(t^2+u^2-6tu)}{(s-m_\gamma^2)} + \frac{(s^2+u^2-6su)}{(t-m_\gamma^2)} + \frac{(t^2+s^2-6ts)}{(u-m_\gamma^2)}\right)\\
&+ i\kappa^2m\epsilon(p_1,p_2,p_3)\left(\frac{t-u}{s(s-m_\gamma^2)} + \frac{t-s}{u(u-m_\gamma^2)} - \frac{s-u}{t(t-m_\gamma^2)}\right).
\end{align}
We have kept the masses the same, but in many cases of physical interest one might wish to consider distinct masses at each vertex, say $m_1$ and $m_2$. It is not hard to do the calculation in this case, and only the first term is modified, giving
\begin{equation}
\cl{M}_4[1,2,3,4]^{(1)} = -\frac{\kappa^2}{8}\left(\frac{4(m_1^2-m_2^2)^2 + t^2+u^2-6tu}{(s-m_\gamma^2)} + s\leftrightarrow u\right).
\end{equation}
Given that there are no massless gravitons in $2+1$ dimensions, you might worry that the massless limit of this amplitude doesn't vanish as expected. You would be right to worry, since in fact we find in the limit that
\begin{equation}
t^2 + u^2 = (t+u)^2 - 2tu = 4(m_1^2+m_2^2)^2 - 2tu,
\end{equation}
such that we end up with
\begin{equation}
\cl{M}_4[1,2,3,4] = -\frac{\kappa^2}{2}\left(\frac{(m_1^2-m_2^2)^2 + (m_1^2 + m_2^2)^2-2tu}{s} + s\leftrightarrow u\right).
\end{equation}
Since all massless particles in $D = 2+1$ are either scalars or fermions, we might imagine that this is a dilaton mediated amplitude that survives the massless limit. This is reminiscent of the vDVZ discontinuity in higher dimensions, where the amplitude for a strictly massless graviton exchange is not the same as for the massless limit of the exchanged massive graviton, due to the contribution of the dilaton that doesn't vanish in the limit. We explore this idea further in appendix. \ref{gravappendix}.

From the perspective of the double copy, this is also not entirely unexpected, since the double copy typically relates spin-1 particles with 
gravitons, dilatons and Kalb-Ramond (axion) fields. Recently it has been demonstrated that the point charge potential in three-dimensions gives rise to a dilaton like potential under the classical double copy \cite{CarrilloGonzalez:2019gof,Gumus:2020hbb}. Furthermore, we also know that scalar-mediated (massless) supergravity amplitudes are non-zero in three dimensions \cite{Bargheer:2012gv}. To compare with this, we can take the all massless limit of the amplitude, taking the external scalars to be part of the super-multiplet, giving rise to the amplitude
\begin{equation}
\cl{A}_4[1,2,3,4] = -\frac{\kappa^2}{2}\left(\frac{t^2+u^2}{s} + \frac{s^2+u^2}{t} + \frac{t^2+s^2}{u}\right).
\end{equation}
This precisely matches the amplitude in the supergravity case \cite{Bargheer:2012gv}, where we have used the fact that e.g. $t = -u$ on shell when $s=0$, and similarly in other channels.
\subsection{Fermion Scattering}
The next simplest four particle amplitude we can consider is two fermions exchanging a gauge boson. This is given by
\begin{align}
\vcenter{\hbox{
		\begin{tikzpicture}[scale=0.5]
		]
		\begin{feynman}  
		\vertex (a) at (-4,2) {$2$};
		\vertex (b) at (-4,-2) {$1$};
		\vertex (c) at (2,-2) {$4$};
		\vertex (d) at (2,2) {$3$};
		\vertex (r) at (0,0);
		\vertex (l) at (-2,0) ;
		\diagram* {
			(a) -- [fermion] (l) -- [photon, momentum={$q$}] (r) -- [fermion] (d),
			(b) -- [fermion] (l) -- [photon] (r) -- [fermion] (c),
		};
		\end{feynman}
		\end{tikzpicture}}} &= \frac{2g^2\braket{12}\braket{\bar{3}\bar{4}}}{\sqrt{p_1\cdot p_2 + m_f^2}\sqrt{p_3\cdot p_4 + m_f^2}(s-m^2)}\left(\frac{x_{12}}{x_{34}}\right)\\
&= -\frac{g^2}{2}\frac{\braket{12}\braket{\bar{3}\bar{4}}\braket{\bar{q}|p_1|\bar{q}}\braket{q|p_4|q}}{2s(s-4m_f^2)(s-m^2)}\\
&= -\frac{g^2}{2}\braket{12}\braket{\bar{3}\bar{4}}\frac{s(t-u) + 4im\epsilon(p_1,p_2,p_3)}{s(s-4m_f^2)(s-m^2)}
\end{align}
where we have used eq. \eqref{fourident}, the fact that $(p_3\cdot p_4 +m_f^2) = -\frac12\left(s-4m_f^2\right)$ and made $1\leftrightarrow 2$ exchange symmetry manifest.

The full amplitude is therefore given by
\begin{equation}
\cl{A}_4 = -\frac{g^2}{2}\braket{12}\braket{\bar{3}\bar{4}}\frac{s(t-u) + 4im\epsilon(p_1,p_2,p_3)}{s(s-4m_f^2)(s-m^2)} -\frac{g^2}{2}\braket{1\bar{3}}\braket{2\bar{4}}\frac{u(t-s) + 4im\epsilon(p_1,p_2,p_3)}{u(u-4m_f^2)(u-m^2)}.
\end{equation}
This amplitude has a strikingly simple form, and is obtained in a far easier manner than by the Feynman rules (compare with appendix \ref{appFermion}). The gravitational version is again obtained via the double copy as

\begin{align}
\cl{M}_4 = &-\frac{\kappa^2}{8}\braket{12}\braket{\bar{3}\bar{4}}\frac{s^2(t^2+u^2-6tu) + 8im(t-u)\epsilon(p_1,p_2,p_3)}{s^2(s-4m_f^2)(s-m^2)}\\ &-\frac{\kappa^2}{8}\braket{1\bar{3}}\braket{2\bar{4}}\frac{u^2(t^2+s^2-6ts) - 8im(t-s)\epsilon(p_1,p_2,p_3)}{u^2(u-4m_f^2)(u-m^2)}.
\end{align}
\subsection{Compton Scattering}
The last example we will focus on is Compton scattering, in this case topologically massive scalar-QED where the $s$-channel amplitude is given by
\begin{align}
\vcenter{\hbox{
		\begin{tikzpicture}[scale=0.5]
		]
		\begin{feynman}  
		\vertex (a) at (-4,2) {$2^+$};
		\vertex (b) at (-4,-2) {$1$};
		\vertex (c) at (2,-2) {$4$};
		\vertex (d) at (2,2) {$3^-$};
		\vertex (r) at (0,0);
		\vertex (l) at (-2,0) ;
		\diagram* {
			(a) -- [photon] (l) -- [plain, momentum={$q$}] (r) -- [photon] (d),
			(b) -- [plain] (l) -- [plain] (r) -- [plain] (c),
		};
		\end{feynman}
		\end{tikzpicture}}} &= \frac{g^2m^2_\phi}{(s-m_\phi^2)}\frac{x_{12}}{x_{34}}\\
&= g^2\frac{\braket{\bar{2}|p_1|\bar{2}}\braket{3|p_4|3}}{4(s-m_\phi^2)m_\gamma^2}
\end{align}

Adding the $u$ channel piece, we find
\begin{equation}
\cl{A}_4[1,2^+,3^-,4] = g^2\frac{\braket{\bar{2}|p_1|\bar{2}}\braket{3|p_4|3}}{(s-m_\phi^2)(u+t-m_\phi^2)} + g^2\frac{\braket{\bar{2}|p_4|\bar{2}}\braket{3|p_1|3}}{(s+t-m_\phi^2)(u-m_\phi^2)}.
\end{equation}
We can immediately repeat the procedure above for gravity by again squaring the $x$-ratio to find in the $s$ channel
\begin{align}
\vcenter{\hbox{
		\begin{tikzpicture}[scale=0.5]
		]
		\begin{feynman}  
		\vertex (a) at (-4,2) {$2^+$};
		\vertex (b) at (-4,-2) {$1$};
		\vertex (c) at (2,-2) {$4$};
		\vertex (d) at (2,2) {$3^-$};
		\vertex (r) at (0,0);
		\vertex (l) at (-2,0) ;
		\diagram* {
			(a) -- [graviton] (l) -- [plain, momentum={$q$}] (r) -- [graviton] (d),
			(b) -- [plain] (l) -- [plain] (r) -- [plain] (c),
		};
		\end{feynman}
		\end{tikzpicture}}} &= \frac{\kappa^2m^4}{(s-m_\phi^2)}\left(\frac{x_{12}}{x_{34}}\right)^2\\
&= \kappa^2\frac{\braket{\bar{2}|p_1|\bar{2}}^2\braket{3|p_4|3}^2}{(s-m_\phi^2)(4m_\gamma^2)^2}
\end{align}
such that
\begin{equation}
\cl{M}_4[1,2^{+2},3^{-2},4] = \kappa^2\frac{\braket{\bar{2}|p_1|\bar{2}}^2\braket{3|p_4|3}^2}{(s-m_\phi^2)(4m_\gamma^2)^2} + \kappa^2\frac{\braket{\bar{2}|p_4|\bar{2}}^2\braket{3|p_1|3}^2}{(u-m_\phi^2)(4m_\gamma^2)^2}.
\end{equation}
While this amplitude is easily reproduced via Feynman diagram, it is possible that there is an additional contact term contribution that is not accounted for. To see this, it would be desireable to have a gauge-dependent polarization vector, in order to perhaps show explicitly that one could gauge-away any contact term contribution. 
In principle, we could derive an on-shell polarization vector that depends on an (possibly light-like) axial-gauge vector, much as one does in four dimensions. We leave this, and the tentative result above, to further study in the future.
\section{The Propagator Double Copy}
Recently, the four dimensional gauge-theory propagator (in the Landau gauge) was shown to enjoy a double copy relationship only if \textit{magnetic} degrees of freedom were included in the propagating spectrum \cite{Moynihan:2020gxj}. Since the magnetic degrees of freedom in three dimensions are in the form of a Chern-Simons term, it seems quite natural from this perspective that topologically massive gauge theory double copies to topologically massive gravity. For freely propagating particles to enjoy this relationship, the propagator ought to double copy. 

Non-Abelian topologically massive gauge theory, or Yang-Mills-Chern-Simons theory, is described by the action \cite{Deser:1981wh}
\begin{equation}
S_{YMCS} =  -\int d^3x~\left[\frac{1}{4e^2}\Tr{F_{\mu\nu}F^{\mu\nu}} - \frac{k}{2}\epsilon^{\mu\nu\rho}\Tr{A_\mu\pd_\nu A_\rho + \frac{2i}{3}A_\mu A_\nu A_\rho}\right],
\end{equation}
where the trace is over gauge group indices, the coupling has mass dimension $[e] = 1/2$ and the Chern-Simons level $k$ is dimensionless. To gauge fix, we can add a gauge-fixing term to the action, of the form
\begin{equation}
\cl{L}_{GF} = -\frac{1}{2\alpha e^2}(\pd_\mu A^\mu)^2.
\end{equation}

Since this part of our discussion will not rely on the particular gauge group under consideration, for simplicity we consider the colour stripped propagator
\begin{equation}\label{propagator}
D_{\mu\nu} = \frac{1}{q^2 + m^2}\left(P_{\mu\nu} -\frac{im\epsilon_{\mu\nu\rho}q^\rho}{q^2}\right) + \alpha\frac{q_\mu q_\nu}{q^4},
\end{equation}
where $P_{\mu\nu} \equiv \eta_{\mu\nu} - \frac{q_\mu q_\nu}{q^2}$.

The graviton propagator is given by
\begin{equation}
	D_{\mu\nu\rho\sigma} = (q^2+m^2)D_{\rho(\mu}D_{\nu)\sigma},
\end{equation}

where we will ignore any terms that vanish when contracted with conserved currents. Working in the Landau gauge where $\alpha = 0$, we find
\begin{align}
2D_{\mu\nu\rho\sigma} &= \frac{1}{q^2+m^2}\left(\eta_{\mu\rho} -\frac{im\epsilon_{\mu\rho\gamma}q^\gamma}{q^2}\right)\left(\eta_{\nu\sigma} -\frac{im\epsilon_{\nu\sigma\lambda}q^\lambda}{q^2}\right) + \mu\leftrightarrow\nu\nn\\
&= \frac{1}{q^2+m^2}\left(\eta_{\mu\rho}\eta_{\nu\sigma} - \frac{m^2}{q^4}\epsilon_{\mu\rho\gamma}\epsilon_{\nu\sigma\lambda}q^\gamma q^\lambda - \frac{im\eta_{\mu\rho}\epsilon_{\nu\sigma\gamma}q^\gamma}{q^2} - \frac{im\eta_{\nu\sigma}\epsilon_{\mu\rho\gamma}q^\gamma}{q^2}\right) + \mu\leftrightarrow\nu\nn\\
&= \frac{1}{q^2+m^2}\left(\eta_{\mu\rho}\eta_{\nu\sigma} + \eta_{\mu\sigma}\eta_{\nu\rho}-\eta_{\mu\nu}\eta_{\sigma\rho} - \frac{im\eta_{\mu\rho}\epsilon_{\nu\sigma\gamma}q^\gamma}{q^2} - \frac{im\eta_{\nu\sigma}\epsilon_{\mu\rho\gamma}q^\gamma}{q^2}\right)\nn\\ &~~~~-\frac{1}{q^2}\left(\eta_{\mu\sigma}\eta_{\nu\rho} - \eta_{\mu\nu}\eta_{\rho\sigma}\right) + \mu\leftrightarrow\nu
\end{align}
where we have used
\begin{equation}
\epsilon_{\mu\nu\gamma}\epsilon_{\rho\sigma\lambda}q^\gamma q^\lambda = -q^2(\eta_{\mu\rho}\eta_{\nu\sigma} - \eta_{\mu\sigma}\eta_{\nu\rho}) +  \cl{O}(q_\mu,q_\nu,q_\rho,q_\sigma),
\end{equation}
where $\cl{O}(q_\mu,q_\nu,q_\rho,q_\sigma)$ vanishes when contracted with conserved currents, and we have split the physical and spurious poles using
\begin{equation}
	\frac{m^2}{q^2(q^2+m^2)} = \frac{1}{q^2} - \frac{1}{q^2+m^2}.
\end{equation}
Putting this all together, we find that the double copy of the propagator is
\begin{align}\label{conserved_gravprop}
D_{\mu\nu,\rho\sigma} = \frac12\bigg(&\frac{\eta_{\mu\rho}\eta_{\nu\sigma} + \eta_{\mu\sigma}\eta_{\nu\rho}-\eta_{\mu\nu}\eta_{\sigma\rho}}{q^2+m^2} - \frac{im\eta_{\rho(\mu}\epsilon_{\nu)\sigma\gamma}q^\gamma}{q^2(q^2+m^2)} - \frac{im\eta_{\sigma(\nu}\epsilon_{\mu)\rho\gamma}q^\gamma}{q^2(q^2+m^2)} \\
&- \frac{\eta_{\mu\rho}\eta_{\nu\sigma} + \eta_{\mu\sigma}\eta_{\nu\rho}-2\eta_{\mu\nu}\eta_{\sigma\rho}}{2q^2}\bigg)
\end{align}

This is precisely the structure of the propagator for topologically massive gravity \cite{Deser:1981wh}, and provides a strong hint that we are on the right track: free topologically massive gauge bosons appear to double copy to free topologically massive gravitons. We note especially that the massless pole comes out naturally, including with the `wrong' sign, such that any massless propagation must arise due to a spin-2 ghost. However, we also see that the coefficient of the massless ghost mode is different when compared with Ref. \cite{Deser:1981wh}. This has no effect on the amplitudes if we consider the case of a definite non-zero mass and everything on-shell (as we did in the above sections), however this does mean that the massless limit becomes more complicated. It is instructive to therefore examine the full propagator, including the pieces that vanish when combined with conserved currents. In this case, we find a  double-copied propagator of the form
\begin{align}
	\cl{D}_{\mu\nu,\rho\sigma} &= \frac{1}{2(q^2+m^2)}\left(P_{\mu\rho} -\frac{im\epsilon_{\mu\rho\gamma}q^\gamma}{q^2}\right)\left(P_{\nu\sigma} -\frac{im\epsilon_{\nu\sigma\lambda}q^\lambda}{q^2}\right) + \mu\leftrightarrow\nu\\
	&=  \frac12\Bigg(-\frac{(q^2-m^2)\left(\eta_{\mu \sigma} \eta_{\nu \rho}+\eta_{\mu \rho} \eta_{\nu \sigma}-2 \eta_{\mu \nu} \eta_{\rho \sigma}\right)}{2q^{2}\left(m^{2}+q^{2}\right)} - \frac{imP_{\rho(\mu}\epsilon_{\nu)\sigma\gamma}q^\gamma}{q^2(q^2+m^2)} - \frac{imP_{\sigma(\nu}\epsilon_{\mu)\rho\gamma}q^\gamma}{q^2(q^2+m^2)}\nn \\
	&+ \frac{2 \eta_{\rho \sigma} \omega_{\mu \nu}-\eta_{\nu \sigma} \omega_{\mu \rho}-\eta_{\mu \sigma} \omega_{\nu \rho}-\eta_{\nu \rho} \omega_{\mu \sigma}-\eta_{\mu \rho} \omega_{\nu \sigma}+2 \eta_{\mu \nu} \omega_{\rho \sigma}+P_{\mu \nu} P_{\rho \sigma}}{m^{2}+q^{2}}\Bigg).\label{gravprop}
\end{align}
This matches the propagator in the de Donder gauge \cite{Pinheiro:1992wv} only if we choose $q^2 = -m^2$ in the numerator of first term in eq. \eqref{gravprop}. However, for generic (off-shell) $q$ this term has residues at both $q^2 = 0$ and $q^2 = -m^2$, which differ by a factor $1/2$ \textit{only} when we leave $q^2$ in the numerator as off-shell. More precisely, we have

\begin{equation}
	\text{Res}\left[\frac{q^2-m^2}{2q^2(q^2+m^2)}\right]_{q^2 = 0} = -\frac12,~~~~~~~~\text{Res}\left[\frac{q^2-m^2}{2q^2(q^2+m^2)}\right]_{q^2 = -m^2} = 1,
\end{equation}
which must be compared with the residue when we take the numerator to be on-shell e.g. $q^2 - m^2 = -2m^2$
\begin{equation}
	\text{Res}\left[\frac{-m^2}{q^2(q^2+m^2)}\right]_{q^2 = 0} = -1,~~~~~~~~\text{Res}\left[\frac{-m^2}{q^2(q^2+m^2)}\right]_{q^2 = -m^2} = 1.
\end{equation}
Taking the numerator to be on-shell then gives the correct TMG double copy, however it may be interesting to examine the double copy where we allow the possibility of a consistent massless limit.
\section{Discussion}
In this paper we have constructed various three dimensional scattering amplitudes in topologically massive gauge theories and gravity. We have constructed the three dimensional equivalent of the $x$-variables, showing that they can be used to directly construct a range of amplitudes for matter-coupled gauge fields or gravitons, and that these agree with those constructed by Feynman diagram. Furthermore, we have motivated the idea that topologically massive gauge and gravity theories in three dimensions are related by the double copy. There are many desirable avenues of study to pursue following this. The most natural next step is to consider pure gauge boson and graviton amplitudes in the context of the double copy and BCJ duality. Since the Feynman rules for Yang-Mills are essentially the same in three and four dimensions, it is likely that BCJ duality at least partially holds for topologically massive theories. It may be that a clever choice of gauge (such as the Gervais-Nueve gauge in three dimensions) might give rise to an off-shell double copy of the vertex as was shown in four dimensions in \cite{Bern:1999ji}. Since we have defined $x$-variables, an obvious next step is to study their deformation, as has been successful in four dimensions \cite{Guevara:2017csg,Guevara:2018wpp,Guevara:2019fsj,Arkani-Hamed:2019ymq,Moynihan:2019bor,Chung:2018kqs,Huang:2019cja,Moynihan:2020gxj}. This will allow us to construct spinning solutions and to study electric-magnetic type dualities, which are extremely rich in 2+1 dimensions, and will be studied elsewhere \cite{Burger:2020}. Another interesting avenue of study is to consider topologically massive supersymmetric theories. It is well known that the superpartner of the graviton | the Rarita-Schwinger field | has a topologically massive construction \cite{Deser:1981wh} and it would be interesting to explore the possibility of a double copy here. Interestingly, this too suffers from a vDVZ-like discontinuity in general \cite{Deser:1977ur}, as was shown recently using amplitudes techniques in four dimensions \cite{Burger:2020kle}, and it would also be interesting to see if this too manifests itself in three dimensions. Finally, there are many interesting quantum aspects of topologically massive gauge theories to explore using on-shell techniques, where it is hopeful that a loop-expansion in topologically massive QCD \cite{Pisarski:1985yj} might be recycled via the double copy to understand loops in topologically massive gravity. We leave these interesting prospects to the future.
\subsection*{Acknowledgements}
I would like to thank Daniel Burger, William Emond, Jeff Murugan and Donal O'Connell for useful discussions and related collaboration.    
NM is supported by funding from the DST/NRF SARChI in Physical Cosmology. Any opinion, finding and conclusion or recommendation expressed in this material is that of the authors and the NRF does not accept any liability in this regard.
\appendix
\section{Conventions and Identities}\label{conventions}
We work in Minkowski space with signature of $(-,+,+)$, with the Mandelstam variables are defined as
\begin{equation}
s = -(p_1+p_2)^2,~~~~~t=-(p_1 + p_3)^2,~~~~~u = -(p_1+p_4)^2.
\end{equation}
The 3D momentum bispinor is given by
\begin{equation}\label{pmat}
p_{\alpha\beta} = p_\mu\sigma^\mu_{\alpha\beta} = \begin{pmatrix}
-p_0 - p_1 & p_2 \\ 
p_2 & -p_0 + p_1
\end{pmatrix}, 
\end{equation}
where $\mu = 0,1,2$, $\det p_{\alpha\beta} = -(-p_0^2+p_1^2+p_2^2) =-m^2$, and the $\sigma$ and $\epsilon$ matrices are given by
\begin{equation}
\sigma^0_{\alpha\beta} = -\begin{pmatrix}
1 & 0 \\ 
0 & 1
\end{pmatrix},~~~~~\sigma^1_{\alpha\beta} = \begin{pmatrix}
-1 & 0 \\ 
0 & 1
\end{pmatrix},~~~~~\sigma^2_{\alpha\beta} = \begin{pmatrix}
0 & 1 \\ 
1 & 0
\end{pmatrix},~~~~~ \epsilon_{\A\B} = \begin{pmatrix}
0 & 1 \\ 
-1 & 0
\end{pmatrix}.
\end{equation}
The gamma matrices are found by raising the last index, i.e. $(\gamma^\mu)_\A^{~\B} = \epsilon^{\beta\gamma}\sigma^\mu_{\A\gamma}$, such that
\begin{equation}
(\gamma^0)_\A^{~\B} = \begin{pmatrix}
0 & 1 \\ 
-1 & 0
\end{pmatrix},~~~~~(\gamma^1)_\A^{~\B} = \begin{pmatrix}
0 & 1 \\ 
1 & 0
\end{pmatrix},~~~~~(\gamma^2)_\A^{~\B} = \begin{pmatrix}
1 & 0 \\ 
0 & -1
\end{pmatrix}.
\end{equation}
These satisfy various trace identities, e.g.
\begin{align}
\Tr{\gamma^\mu\gamma^\nu} &= 2\eta^{\mu\nu},~~~~~\Tr{\gamma^\mu\gamma^\nu\gamma^\rho} = 2\epsilon^{\mu\nu\rho},\\\Tr{\gamma^\mu\gamma^\nu\gamma^\rho\gamma^\sigma} &= 2\left(\eta^{\mu\nu}\eta^{\rho\sigma} - \eta^{\mu\rho}\eta^{\nu\sigma} + \eta^{\mu\sigma}\eta^{\rho\nu}\right),
\end{align} 
as well as the algebra
\begin{align}\label{gammaalgebra}
\gamma^\mu\gamma^\nu &= \eta^{\mu\nu}\mathds{1} + \epsilon^{\mu\nu\rho}\gamma_\rho,\\
\gamma^\mu\gamma^\nu\gamma^\rho &= \gamma^{[\mu}\gamma^\nu\gamma^{\rho]} + \eta^{\mu\rho}\gamma^\nu + \eta^{\rho\nu}\gamma^\mu - \eta^{\mu\nu}\gamma^\rho,\\
&= \epsilon^{\mu\nu\rho} + \eta^{\mu\rho}\gamma^\nu + \eta^{\rho\nu}\gamma^\mu - \eta^{\mu\nu}\gamma^\rho,
\end{align}
the completeness relation
\begin{equation}
(\gamma^\mu)_{\alpha}^{~\beta}(\gamma_\mu)_{\rho}^{~\sigma} = -\left(\epsilon_{\alpha\rho}\epsilon^{\beta\sigma} + \delta^{\sigma}_\alpha\delta^{\beta}_\rho\right),
\end{equation}
and the Levi-Civita relation
\begin{equation}\label{leviCID}
\epsilon^{\rho\mu\nu} q_\rho\braket{i|\gamma_\mu|j}\braket{k|\gamma_\nu|l} = 2\braket{k|q|j}\braket{il} - \braket{i|q|j}\braket{kl} - \braket{k|q|l}\braket{ij}.
\end{equation}
Using these, we can derive the spinor-helicity identities
\begin{align}\label{fourident}
\braket{a|p_1p_2p_3|b} &= \braket{ab}\epsilon(p_1,p_2,p_3) + 2(p_1\cdot p_2)\braket{a|p_3|b} - (p_1\cdot p_3)\braket{a|p_2|b}\\
\braket{a|p_1p_2|b} &= \braket{ab}(p_1\cdot p_2) + \epsilon^{\mu\nu\rho}p_{1\mu}p_{2\nu}\braket{a|\gamma_\rho|b}\nn
\end{align}
For $b = \bar{a}$, this simplifies to become
\begin{align}\label{fourident}
\braket{a|p_1p_2p_3|\bar{a}} &= -2im\epsilon(p_1,p_2,p_3) + 4(p_1\cdot p_2)(p_3\cdot p_a) - 2(p_1\cdot p_3)(p_2\cdot p_a)\\
\braket{a|p_1p_2|\bar{a}} &= -2im(p_1\cdot p_2) + 2\epsilon^{\mu\nu\rho}p_{1\mu}p_{2\nu}p_{a\rho}.
\end{align}
We decompose the bispinor as $p_{\alpha\beta} = \lambda_{(\A}\bar{\lambda}_{\B)}$, where the spinors $\lambda,\bar{\lambda}$ are given by
\begin{equation}
\lambda_\A = \frac{1}{\sqrt{p_0-p_1}}\begin{pmatrix}
p_2 -im\\
p_1 - p_0 
\end{pmatrix},~~~~~\bar{\lambda}_\A = -\frac{1}{\sqrt{p_0-p_1}}\begin{pmatrix}
p_2 +im\\
p_1 - p_0 
\end{pmatrix}.
\end{equation}
We note that these are related to the conventional 2+1 dimensional spinors by $$u(p_i) = \bra{i} = -\bar{v}(p_i),~~~\bar{u}(p_i) = \ket{\bar{i}} = -v_{p_i}$$

The Dirac equations are given by
\begin{equation}
p_j\ket{j} = -im_j\ket{j},~~~~~p_j\ket{\bar{j}} = im_j\ket{\bar{j}},~~~~~\bra{j}p_j = im_j\bra{j},~~~~~\bra{\bar{j}}p_j = -im_j\bra{\bar{j}}
\end{equation}

These forms allow us to write 
\begin{equation}\label{unsym1}
\lambda_\A\bar{\lambda}_\B = p_{\A\B} + im\epsilon_{\A\B},~~~~~\bar{\lambda}_\A\lambda_\B = p_{\A\B} - im\epsilon_{\A\B},
\end{equation}
which immediately implies that $\lambda_{(\A}\bar{\lambda}_{\B)} = p_{\alpha\beta}$ and that
\begin{align}
\epsilon^{\alpha\beta}\lambda_{\beta}\bar{\lambda}_{\alpha}= \braket{\lambda\bar{\lambda}} &= -\braket{\bar{\lambda}\lambda} = -2im.
\end{align}
We can contract in a $\sigma$ matrix to find that $\lambda_\alpha\bar{\lambda}_\beta\sigma^{\mu\alpha\beta} = p_{\alpha\beta}\sigma^{\mu\alpha\beta} + im\Tr{\gamma^\mu}$ and therefore that
\begin{equation}
\braket{i|\gamma^\mu|\bar{i}} = \braket{\bar{i}|\gamma^\mu|i} = 2p^\mu.
\end{equation}
Using the above relations we can derive the following identities
\begin{equation}
\braket{i\bar{j}}\braket{\bar{i}j} = s_{ij} - (m_i-m_j)^2,~~~~~~\braket{\bar{i}\bar{j}}\braket{ij} = s_{ij} - (m_i+m_j)^2,
\end{equation}
with $s_{ij} = -(p_i+p_j)^2$.

For four same-mass particles $i,j,k,l$ that share a propagator, we have the identity
\begin{equation}
\Tr{p_i p_j p_k} = 2\epsilon^{\mu\nu\rho}p_{i\mu}p_{j\nu}p_{k\rho} = i\sqrt{s_{ij}s_{jk}s_{kl}} = \braket{\bar{i}j}\braket{\bar{j}k}\braket{\bar{k}i}.
\end{equation}

For four particle scattering with momentum conservation dictated by $\sum_{i=1}^4 p_i = 0$, the following identities are often useful \cite{Agarwal:2008pu}

\begin{equation}\label{ratioids}
\frac{\braket{lm}}{\braket{kn}} = \frac{\braket{\bar{k}l}}{\braket{\bar{m}n}} = \frac{\braket{\bar{k}\bar{n}}}{\braket{\bar{l}\bar{m}}},~~~~~\{k,l,m,n\} = \{1,2,3,4\}.
\end{equation}
Throughout this text, we use the Levi-Civita \textit{tensor} in Minkowski space, defined by $\epsilon^{012} = \epsilon_{012} = +1$, with the relationship between upper and lower given by
\begin{equation}
\epsilon^{\alpha\beta\gamma}\,\eta_{\alpha\kappa}\,\eta_{\beta\lambda}\,\eta_{\gamma\mu} \,=\, \epsilon_{\kappa\lambda\mu},
\end{equation}
 and many identities used in the main text can be derived from the relation
\begin{equation}
\epsilon^{\mu\nu\rho}\epsilon_{\alpha\beta\gamma} = -\begin{vmatrix}
\delta^\mu_\alpha & \delta^\mu_\beta & \delta^\mu_\gamma \\ 
\delta^\nu_\alpha & \delta^\nu_\beta & \delta^\nu_\gamma  \\ 
\delta^\rho_\alpha & \delta^\rho_\beta & \delta^\rho_\gamma
\end{vmatrix}. 
\end{equation}
We note that the relationship between the tensor $\epsilon$ and the symbol $\varepsilon$ is, given our conventions,
\begin{equation}
\epsilon^{\mu\nu\rho} = \varepsilon^{\mu\nu\rho},~~~~~\epsilon_{\mu\nu\rho} = -\varepsilon_{\mu\nu\rho}.
\end{equation}
We make frequent use of the Schouten identity, which is given by
\begin{equation}
	\epsilon_{\A\B}\epsilon_{\gamma\delta} + \epsilon_{\A\gamma}\epsilon_{\delta\B} + \epsilon_{\A\delta}\epsilon_{\B\gamma} = 0.
\end{equation}
It is often more useful in the spinor-helicity form, where we simply contract in four arbitrary spinors to find a relationship of the form
\begin{equation}
	\braket{ij}\braket{kl} + \braket{ik}\braket{lj} + \braket{il}\braket{jk} = 0, 
\end{equation}
where bars can be freely added as needed.
\section{Feynman Rules}
Here we collect the Feynman rules used to compute some quantities in this appendix.

The colour stripped topologically massive gauge-boson propagator is given by
\begin{equation}\label{gbpropagator}
D_{\mu\nu} = \frac{1}{q^2 - k^2e^4}\left(P_{\mu\nu} -\frac{ike^2\epsilon_{\mu\nu\rho}q^\rho}{q^2}\right),
\end{equation}
where $P_{\mu\nu} = \eta_{\mu\nu} - \frac{q_\mu q_\nu}{q^2}$.

The scalar-scalar-gauge boson and fermion-fermion-gauge boson vertices are given by
\begin{equation}
V_{scalar}^\mu = -ig(p_1^\mu - p_2^\mu),~~~~~V_{fermion}^\mu = -ig\braket{1|\gamma^\mu|2},
\end{equation}
while the massive graviton propagator is given by
\begin{equation}\label{massivegravprop}
\cl{D}(q)_{\mu\nu\rho\sigma} = \frac{i}{2q^2}\left(P_{\mu\rho}P_{\nu\sigma}+P_{\mu\sigma}P_{\nu\rho} - \frac{2}{D-1}P_{\mu\nu}P_{\rho\sigma}\right).
\end{equation}
The scalar-scalar-graviton vertex is given by
\begin{equation}\label{scalarvertex}
V^{\mu\nu} = -i\frac{\kappa}{2}\left(p_1^\mu p_2^\nu+p_1^\nu p_2^\mu - \frac{1}{2}\eta^{\mu\nu} q^2\right)
\end{equation}
\section{Gauge-Theory Three-Particle Amplitude}
We can derive an arbitrary spin three particle on-shell amplitude from the Feynman rules by contracting polarization vectors with the vertex to find
\begin{align}
A[1^-,2^-,3^+] = \frac{if^{abc}}{8m_am_bm_c}\bigg(&i\mu\braket{12}\braket{2\bar{3}}\braket{\bar{3}1} + \braket{\bar{3}|p_1-p_2|\bar{3}}\braket{12}^2\\& + \braket{2|p_3-p_1|2}\braket{1\bar{3}}^2 + \braket{1|p_2-p_3|1}\braket{\bar{3}2}^2\bigg),\nn
\end{align}
where all other spin configurations are found by adding or removing bars. This can be greatly simplified by considering e.g.
\begin{align*}
\braket{1|p_2-p_3|1}\braket{\bar{3}2}^2 &= \braket{12}\braket{\bar{2}1}\braket{\bar{3}2}^2 - \braket{13}\braket{\bar{3}1}\braket{\bar{3}2}^2\\
&= -\braket{12}\braket{\bar{3}2}\left(\braket{\bar{2}\bar{3}}\braket{21} + \braket{\bar{2}2}\braket{1\bar{3}}\right) -\braket{1\bar{3}}\braket{\bar{3}2}\left(\braket{1\bar{3}}\braket{23} + \braket{12}\braket{3\bar{3}}\right)\nn\\
&= -\braket{12}\braket{2\bar{3}}\braket{\bar{3}1}\braket{\bar{2}2} + \braket{\bar{3}|p_2|\bar{3}}\braket{12}^2 - \braket{12}\braket{2\bar{3}}\braket{\bar{3}1}\braket{3\bar{3}} -\braket{2|p_3|2}\braket{1\bar{3}}^2\\
&= 2i\braket{12}\braket{2\bar{3}}\braket{\bar{3}1}\left(m_2-m_3\right) + \braket{\bar{3}|p_2|\bar{3}}\braket{12}^2 -\braket{2|p_3|2}\braket{1\bar{3}}^2.
\end{align*}

Similarly, we have
\begin{align}
\braket{2|p_3-p_1|2}\braket{\bar{3}2}^2 &= 2i\braket{12}\braket{2\bar{3}}\braket{\bar{3}1}\left(m_3-m_1\right) + \braket{1|p_3|1}\braket{2\bar{3}}^2 -\braket{\bar{3}|p_1|\bar{3}}\braket{12}^2\\
\braket{\bar{3}|p_1-p_2|\bar{3}}\braket{12}^2 &= 2i\braket{12}\braket{2\bar{3}}\braket{\bar{3}1}\left(m_1-m_2\right) + \braket{2|p_1|2}\braket{1\bar{3}}^2 -\braket{1|p_2|1}\braket{2\bar{3}}^2
\end{align}

Taking everything to have the same (topological) mass $m$, we find
\begin{equation}\label{red3pt}
A[1^-,2^-,3^+] = \frac{if^{abc}}{8m^3}\left(im\braket{12}\braket{2\bar{3}}\braket{\bar{3}1} + \braket{\bar{3}|p_1|\bar{3}}\braket{12}^2 + \braket{2|p_3|2}\braket{1\bar{3}}^2\right).
\end{equation}
\section{Gordon Identities}
In this section we derive the Gordon identities in spinor-helicity notation. First, using the Dirac equation, we can write
\begin{align}
\braket{i|\gamma^\mu|j} &= i\frac{\braket{i|\gamma^\mu\gamma^\nu|j}p_{j\nu}}{2m_j} - i\frac{\braket{i|\gamma^\nu\gamma^\mu|j}p_{i\nu}}{2m_i}\\
&= i\frac{\braket{i|\eta^{\mu\nu} + \epsilon^{\mu\nu\rho}\gamma_\rho|j}p_{j\mu}}{2m_j} - i\frac{\braket{i|\eta^{\mu\nu} - \epsilon^{\mu\nu\rho}\gamma_\rho|j}p_{i\mu}}{2m_i}\\
&= \frac{i}{2}\left[\left(\frac{p_j^\mu}{m_j} - \frac{p_i^\mu}{m_i}\right)\braket{ij} + \left(\frac{p_{j\nu}}{m_j} - \frac{p_{i\nu}}{m_i}\right)\epsilon^{\mu\nu\rho}\braket{i|\gamma_\rho|j}\right]
\end{align}
For $m_i = m_j = m$, we can write this in the more familiar form
\begin{align}
\braket{i|\gamma^\mu|j} = -\frac{i}{2m}\left[(p_i^\mu - p_j^\mu)\braket{ij} + \left(p_{i\nu} + p_{j\nu}\right)\epsilon^{\mu\nu\rho}\braket{i|\gamma_\rho|j}\right].
\end{align}
Similarly, we can do this for a single barred spinor to find
\begin{align}
\braket{\bar{i}|\gamma^\mu|j} &= -\frac{i}{2m}\left[(p_i^\mu + p_j^\mu)\braket{\bar{i}j} + \left(p_{i\nu} - p_{j\nu}\right)\epsilon^{\mu\nu\rho}\braket{\bar{i}|\gamma_\rho|j}\right],
\end{align}
with the other possibilities related by complex conjugation.

We can also derive the identities
\begin{align}
\braket{i|\gamma^\mu|j} &= \frac{\braket{i|\gamma^\nu\gamma^\mu\gamma^\rho|j}p_{i\nu}p_{j\rho}}{m^2}\\
&= \frac{\braket{i|p_j|j}p_{i}^\mu - \epsilon^{\mu\nu\rho}p_{i\nu}p_{j\rho}\braket{ij} + (\delta^\mu_\rho\delta^\nu_\sigma - \delta^\nu_\rho\delta^\mu_\sigma)\braket{i|\gamma^\sigma|j}p_{i\nu}p_j^\rho}{m^2}\\
&= -\frac{im(p_{i}^\mu-p_j^\mu)\braket{ij} + \epsilon^{\mu\nu\rho}p_{i\nu}p_{j\rho}\braket{ij} + \braket{i|\gamma^\mu|j}p_{i}\cdot p_j}{m^2}\\
&= -\braket{ij}\frac{im(p_{i}^\mu-p_j^\mu) + \epsilon^{\mu\nu\rho}p_{i\nu}p_{j\rho}}{m^2 + p_i\cdot p_j},
\end{align}
and similarly
\begin{equation}
\braket{i|\gamma^\mu|\bar{j}} =  \braket{i\bar{j}}\frac{im(p_{i}^\mu+p_j^\mu) + \epsilon^{\mu\nu\rho}p_{i\nu}p_{j\rho}}{m^2 - p_i\cdot p_j}.
\end{equation}
\section{A Selection of Amplitudes via the Feynman Rules}\label{feynappendix}
For comparison purposes, in this appendix we compute some of the amplitudes by Feyman diagram.
\subsection{Scalar Scattering via Gauge Boson Exchange}
The $2\rightarrow 2$ $s$-channel scattering of scalars via a gauge boson is given by
\begin{align}
\cl{A}_4[1234] &= V^\mu(p_1,p_2) D_{\mu\nu}V^\nu(p_3,p_4) = g(p_1-p_2)^\mu D_{\mu\nu}g(p_3-p_4)^\nu\\
&= g^2\frac{-m^2(p_1-p_2)\cdot(p_3-p_4) + im\epsilon_{\mu\nu\rho}(p_1-p_2)^\mu(p_3-p_4)^\nu q^\rho}{q^2(q^2 + m^2)},
\end{align}
For the Maxwell propagator, starting with the $t$ channel, this is given by\footnote{Note that $V^\mu q_\mu = 0$ in the same mass case, since $V^\mu q_\mu = (p_1-p_2)\cdot (p_1+p_2) = -(m_1^2 - m_2^2)$ in general.}
\begin{align}
\cl{A}_4[1,2,3,4] &= V(1,2)^\mu\frac{\eta_{\mu\nu}}{t+m^2}V(3,4)^\nu + 2\leftrightarrow 3 + 2\leftrightarrow 4\\
&= \frac{2e^2}{t+m^2}\left(p_1\cdot p_3 - p_1\cdot p_4\right) + 2\leftrightarrow 3 + 2\leftrightarrow 4\\
&= e^2\left(\frac{u-s}{t+m^2} + \frac{t-s}{u+m^2} + \frac{u-t}{s+m^2}\right).\label{m4pt}
\end{align}
For the Chern-Simons propagator, this is
\begin{align}
\cl{A}_4[1,2,3,4] &= V(1,2)^\mu\frac{im\epsilon_{\mu\nu\rho}q^\rho}{t(t+m^2)}V(3,4)^\nu + 2\leftrightarrow 3 + 2\leftrightarrow 4\\
&= 4ie^2m\epsilon(123)\left(\frac{1}{s(s+m^2)}+\frac{1}{u(u+m^2)}-\frac{1}{t(t+m^2)}\right)\\
&= 2ie^2m\braket{\bar{1}2}\braket{\bar{2}3}\braket{\bar{3}1}\left(\frac{1}{s(s+m^2)}+\frac{1}{u(u+m^2)}-\frac{1}{t(t+m^2)}\right)
\end{align}
where we have used the notation $\epsilon(ijk) = \epsilon_{\mu\nu\rho}p_i^\mu p_j^\nu p_k^\rho$. In order to express this in a simple form, we used the fact that, in the case of the $t$-channel,
\begin{equation}
\epsilon(13q) = \epsilon(1(q-4)q) = -\epsilon(14q),
\end{equation} 
and similar simplifications in other channels.
\subsection{Scalar Scattering via Graviton Exchange}
To compute the same thing in the gravitational theory, we compute the amplitude
\begin{align}
\cl{A}_4[1234] &= V^{\mu\nu}(p_1,p_2) D_{\mu\nu\rho\sigma}V^{\rho\sigma}(p_3,p_4) 
\\&= \kappa^2\frac{(p_1^\mu p_2^\nu + p_1^\mu p_2^\nu - \frac12q^2\eta^{\mu\nu})\left(I_{\mu\nu\rho\sigma} -iJ_{\mu\nu\rho\sigma}\right)(p_3^\mu p_4^\nu + p_3^\mu p_4^\nu - \frac12q^2\eta^{\mu\nu})}{2(s^2-m^2)},
\end{align}
where 
\begin{align}
I_{\mu\nu\rho\sigma} &= \eta_{\mu\nu}\eta_{\rho\sigma} + \eta_{\mu\sigma}\eta_{\nu\rho} - \eta_{\mu\rho}\eta_{\nu\sigma},\\ J_{\mu\nu\rho\sigma} &= \frac{mP_{\mu\nu}\epsilon_{\rho\sigma\lambda}q^\lambda}{2q^2} + \frac{mP_{\rho\sigma}\epsilon_{\mu\nu\lambda}q^\lambda}{2q^2} +\frac{mP_{\rho\nu}\epsilon_{\mu\sigma\lambda}q^\lambda}{2q^2} + \frac{mP_{\mu\sigma}\epsilon_{\rho\nu\lambda}q^\lambda}{2q^2},
\end{align}
and $P_{\mu\nu} = \eta_{\mu\nu} - \frac{q_\mu q_\nu}{q^2}$. We note that the other term in the propagator in eq. \eqref{gravprop} does not contribute when coupled to conserved currents, and in fact, neither does $\frac{q_\mu q_\nu}{q^2}$. With the aid of FeynCalc \cite{Shtabovenko:2020gxv}, we find the amplitude in the $s$ channel to be
\begin{equation}
\cl{A}_4[1234] = \frac{\kappa^2}{32}\frac{4(m_1^2-m_2^2)^2 - (t^2+u^2-6tu)}{s-m^2} + im\frac{\kappa^2}{4} \frac{(t-u)\epsilon(p_1,p_2,p_3)}{s(s-m_\gamma^2)}.
\end{equation}
\subsection{Fermion Scattering via Gauge Boson Exchange}\label{appFermion}
For comparison, lets compute the Fermion amplitude directly from the Feynman rules. The $s$ channel diagram is given by
\begin{align}
\cl{A}_4[1234] &= V^\mu(p_1,p_2) D_{\mu\nu}V^\nu(p_3,p_4) = \braket{\bar{1}|\gamma^\mu|\bar{2}}D_{\mu\nu}\braket{3|\gamma^\nu|4}\\
&= \frac{-m^2\braket{\bar{1}|\gamma^\mu|\bar{2}}\eta_{\mu\nu}\braket{3|\gamma^\nu|4} + im\epsilon_{\mu\nu\rho}\braket{\bar{1}|\gamma^\mu|\bar{2}}\braket{3|\gamma^\nu|4}q^\rho}{q^2(q^2 + m^2)},
\end{align}
where we have used $\braket{\bar{1}|q|\bar{2}} = 0$ for $m_1 = m_2$, and similarly for $3,4$.

Let's do this a piece at a time. The first part, using the Gordon identities, gives
\begin{align}
\braket{\bar{1}|\gamma^\mu|\bar{2}}\braket{3|\gamma_\mu|4} &= \frac{\braket{\bar{1}\bar{2}}\braket{34}}{(m_f^2 + p_1\cdot p_2)(m_f^2 + p_3\cdot p_4)}\left[(im_f(p_1-p_2)^\mu + \epsilon^{\mu\nu\rho}p_{1\nu}p_{2\rho})(im_f(p_3-p_4)_\mu - \epsilon_{\mu\alpha\beta}p_3^\alpha p_{4}^\beta)\right]\\
&= \frac{\braket{\bar{1}\bar{2}}\braket{34}}{(m_f^2 + p_1\cdot p_2)(m_f^2 + p_3\cdot p_4)}\left[m_f^2(u-t) +\left(\delta^\nu_\alpha\delta^\rho_\beta - \delta^\nu_\beta\delta^\rho_\alpha\right)p_{1\nu}p_{2\rho}p_3^\alpha p_{4}^\beta\right]\\
&= -\frac{\braket{\bar{1}\bar{2}}\braket{34}(t-u)(s-4m_f^2)}{4(s-4m_f^2)^2}\\
&= -\frac{\braket{\bar{1}\bar{2}}\braket{34}(t-u)}{4(s-4m_f^2)}
\end{align}
where we have used $p_1\cdot p_2 + m_f^2 = -\frac12s + 2m_f^2 = \frac12(t+u)$.

The second piece is given by\footnote{To see this we note that we can express $$\epsilon_{\mu\nu\rho}p_2^\mu p_3^\nu q^\rho = \frac14\epsilon_{\mu\nu\rho}\braket{\bar{2}|\gamma^\mu|2}\braket{\bar{3}|\gamma^\nu|3} q^\rho$$
using $p_i^\mu = \frac12\braket{i|\gamma^\mu|\bar{i}}$ and then use the identity in eq. \eqref{leviCID}.}
\begin{align}
\epsilon_{\mu\nu\rho}\braket{\bar{1}|\gamma^\mu|\bar{2}}\braket{3|\gamma^\nu|4}q^\rho &= -2\braket{3|q|\bar{2}}\braket{\bar{1}4}\\
&= -4\epsilon_{\mu\nu\rho}p_2^\mu p_3^\nu q^\rho\frac{\braket{\bar{1}4}}{\braket{2\bar{3}}} -im(p_2+p_3)\cdot q\frac{\braket{\bar{1}4}}{\braket{2\bar{3}}}
\\&= 4\epsilon_{\mu\nu\rho}p_1^\mu p_2^\nu p_3^\rho\frac{\braket{\bar{1}\bar{2}}}{\braket{\bar{3}\bar{4}}} \\&= -4\epsilon_{\mu\nu\rho}p_1^\mu p_2^\nu p_3^\rho\frac{\braket{\bar{1}\bar{2}}\braket{34}}{s-4m_f^2}
\end{align}
where we have used $(p_2+p_3)\cdot q = (p_1\cdot p_2 + p_1\cdot p_3 + p_1\cdot p_4 -m_f^2) = -\frac12(s+t+u) + 2m_f^2 = 0$ and, on the last line, eq. \eqref{ratioids} and $\braket{34}\braket{\bar{3}\bar{4}} = -2(p_3\cdot p_4 +m_f^2) = s-4m_f^2$. Putting this all together gives the amplitude
\begin{equation}
\cl{A}_4[1^{+1/2}2^{+1/2}3^{-1/2}4^{-1/2}] = \frac{\braket{\bar{1}\bar{2}}\braket{34}(t-u) -4im\braket{\bar{1}\bar{2}}\braket{34}\epsilon_{\mu\nu\rho}p_1^\mu p_2^\nu p_3^\rho}{2(s-4m_f^2)(s-m^2)}
\end{equation}
\section{Polarization Vectors}
Topologically massive theories have the odd property that they are both massive \textit{and} gauge invariant, unlike theories endowed with a generic Proca mass which lack any gauge freedom. This means that constructing polarization vectors isn't straightforward \cite{Kogan:1985qn,Pisarski:1985yj,Kogan:1990xg,Banerjee:2000gc}, since one needs to consider encoding both the massive physical degrees of freedom and the gauge degrees of freedom into one object. On shell, this is slightly less problematic, and we know what we need to satisfy: we need a transverse polarization vector with $\pm1$ spin degrees of freedom and a gauge choice that can result in no more than a total derivative in the action. The tensor product of two vectors with different helicities should also produce the tensorial structure of the off-shell propagator. That all being said, constructing such an object is not entirely without its difficulties, and so it will suffice for our purposes to find a specific gauge-fixed polarization vector.

The equations of motion for the gauge field $A_\mu^a$ are given by
\begin{equation}
\left(\eta_{\mu\rho}\pd^\nu + \frac{m}{2}\epsilon_{\mu\nu\rho}\right)F^{\nu\rho} = 0.
\end{equation}
In momentum space, we can find a plane-wave solution by assuming the form $A_\mu^a(q) = c^a\epsilon^\mu(q) e^{iq\cdot x}$, which we need to satisfy the equation
\begin{equation}\label{landauEOM}
\left(-q^2\eta_{\mu\rho} + q_\mu q_\rho + im\epsilon_{\mu\nu\rho}q^\nu\right)\epsilon^\rho(q)e^{iq\cdot x} = 0.
\end{equation}
We need to choose a gauge at this point, and we will choose the Landau gauge $q_\mu \epsilon^\mu(q) = 0$, which reduces the problem to finding a vector that satisfies
\begin{equation}\label{landauEOM}
q^2\epsilon^\mu(q) = im\epsilon^{\mu\nu\rho}q_\nu\epsilon_\rho(q).
\end{equation}
Since we are working in spinor-helicity variables, we will make an Ansatz of the form $\epsilon^\mu = \braket{a|\gamma^\mu|b}$ and fix $a,b$ such that it solves the equation and is correctly normalised. To this end, we can use the Poincar\'e algebra to write the right hand side as
\begin{align}
imq_\nu\braket{a|\epsilon^{\mu\nu\rho}\gamma_\rho|b} &= imq_\nu\braket{a|\gamma^\mu\gamma^\nu - \eta^{\mu\nu}|b}\\ &= im\left(\braket{a|\gamma^\mu q|b} - q^\mu\braket{ab}\right)
\end{align}
Choosing $a = b = q$ then gives 
\begin{equation}
im\braket{q|\gamma^\mu q|q} = \frac{im}{2}\braket{\bar{q}q}\epsilon^\mu = q^2\epsilon^\mu,
\end{equation}
which is precisely the left hand side of eq. \eqref{landauEOM}, where we have used $\braket{\bar{q}q} = 2im$ and $m^2 = -q^2$ on-shell.

Normalising this such that $\epsilon^\mu\epsilon_\mu^* = -1$, we find polarization vectors of the form
\begin{equation}
\epsilon^{\mu-} = \frac{\braket{q|\gamma^\mu|q}}{2|q|},~~~~~\epsilon^{\mu+} = \frac{\braket{\bar{q}|\gamma^\mu|\bar{q}}}{2|q|}.
\end{equation}
One can verify that these polarization vectors reproduce the Landau gauge propagator given by eq. \eqref{gbpropagator}, i.e. that
\begin{equation}
\Delta_{\mu\nu} = \epsilon^{-}_\mu\epsilon^+_{\nu}
\end{equation}

We also note that these polarization vectors can also be used for pure massive Yang-Mills, where we now sum over two possible modes to find
\begin{equation}
\Delta^{YM}_{\mu\nu} = \frac12\sum_{h = \pm}\epsilon^{-h}_\mu\epsilon^{h}_{\nu} = \eta_{\mu\nu} - \frac{q_\mu q_\nu}{q^2}.
\end{equation}
The polarization vector of a massless Yang-Mills field is given in \cite{Lipstein:2012kd} as
\begin{equation}
\epsilon^{\mu} = \frac{\braket{q|\gamma^\mu|\eta}}{\braket{q\eta}}.
\end{equation}
This could in principle be used to construct $x$-variables for massless fields coupled to conserved currents, which would then be of the form
\begin{equation}
x_1 = \frac{\braket{q|u_1|\eta}}{\braket{q\eta}}.
\end{equation}
\section{Graviton vs Dilaton Exchange in General Dimensions}\label{gravappendix}
For a gravitational theory in $D$ dimensions, $[\kappa^2_D] = [G_D] = 2-D$. However, the graviton has \textit{massless} degrees of freedom $\frac{D(D-3)}{2}$ and \textit{massive} d.o.f $\frac{D(D-1)}{2} - 1$ \cite{Hinterbichler:2011tt}. Thus, for $D = 3$ we find that $[G] = -1$, that there are no massless gravitons and that massive gravitons have two degrees of freedom. In general dimensions, the massless dilaton polarization tensor is of the form 
\begin{equation}
\epsilon^{d}(q)_{\mu\nu} = \frac{1}{\sqrt{D-2}}\left(\eta_{\mu\nu} - q_\mu\eta_\nu - q_\nu\eta_\mu\right),
\end{equation}
where $\eta_\mu$ is an auxiliary vector that satisfies $\eta\cdot q = 1$ and $\eta^2 = 0$. For a massive dilaton, the polarization tensor is given by
\begin{equation}
\cl{E}^{d}(q)_{\mu\nu} = \frac{1}{\sqrt{D-1}}\left(\eta_{\mu\nu} - \frac{q_\mu q_\nu}{q^2}\right).
\end{equation}
In this section, we will compare dilaton exchange amplitudes with actual graviton exchange amplitudes in general dimensions.

Two scalars exchanging a massless graviton in $D$-dimensional general relativity is given by the amplitude
\begin{align}
\cl{M}_4[1,2,3,4] &= V^{\mu\nu}(p_1,p_2)D_{\mu\nu\rho\sigma}V^{\rho\sigma}(p_3,p_4)\\
&= \frac{i \kappa_D^2 \left((D-2)m_1^4+m_1^2 \left(D \left(2m_2^2-t-u\right)+2 (t+u)\right)+(D-2) \left(m_2^2-t\right) \left(m_2^2-u\right)\right)}{4 (D-2) s},\nn
\end{align}
where the vertices and propagator are given by eqs. \eqref{scalarvertex} and \eqref{gravprop} respectively.

Taking the non-relativistic limit here gives an amplitude that vanishes in $D=3$ and is divergent in $D=2$
\begin{equation}
\cl{M}_4[1,2,3,4]\bigg|^{NR} = -i\kappa_D^2\frac{(D-3) m_1^2 m_2^2}{(D-2) s},
\end{equation}
 implying that there is no Newtonian potential in $D\leq 4$ General Relativity as expected. If we consider a massless dilaton in the spectrum, this is not the case, since we now have a dilaton exchange amplitude of the form
\begin{align}
\cl{M}_4[1,2,3,4]\bigg|_{dilaton} &= \frac{(V^{\mu\nu}(p_1,p_2)\epsilon^d_{\mu\nu}(q))(V^{\rho\sigma}(p_3,p_4)\epsilon^d_{\rho\sigma}(q))}{s}\\
&= \frac{i \kappa_D ^2 \left((D-2) s+4 m_1^2\right) \left((D-2) s+4 m_2^2\right)}{16 (D-2) s}, 
\end{align}
which gives a non-relativistic amplitude
\begin{equation}
\cl{M}_4[1,2,3,4]\bigg|_{dilaton}^{NR} = \frac{i \kappa_D ^2 m_1^2 m_2^2}{(D-2) s},
\end{equation}
implying that the dilaton will produce a classical Newtonian potential in $D\geq 2$. 

For a massive gravity in $D$ dimensions, we find an amplitude of the form
\begin{align}
\cl{M}_4[1,2,3,4] &= V^{\mu\nu}(p_1,p_2)\cl{D}_{\mu\nu\rho\sigma}V^{\rho\sigma}(p_3,p_4)\\
&= -\frac{i \kappa_D ^2 \left((D-2) t^2-2 D t u+(D-2) u^2+4 \left(m_1^2-m_2^2\right)^2\right)}{16 (D-1) \left(M^2-s\right)}
\end{align}
which has a non-relativistic limit
\begin{equation}
\cl{M}_4[1,2,3,4]\bigg|^{NR} = \frac{i\kappa_D ^2 (D-2)  m_1^2 m_2^2}{(D-1) \left(s-M^2\right)}
\end{equation}
We see then that the massive graviton does have a Newtonian like classical limit in $D = 3$, unlike the massless graviton. Furthermore, we notice that this amplitude is non-zero even if we take $D\rightarrow 3$ and the mass of the graviton to zero | a manifestation of the well known vDVZ discontinuity. In fact, we see that taking $M\rightarrow 0$ gives rise to an overall factor difference of $(D-3)/(D-2)$ vs $(D-2)/(D-1)$ when comparing the strictly massless graviton and the massless limit of the massive graviton, which reproduces the four-dimensional factor of $2/3$ \cite{vanDam:1970vg,Moynihan:2017tva}.

Equally, we find that the massive dilaton will similarly produce a potential, since the massive dilaton exchange amplitude is given by
\begin{align}
\cl{M}_4[1,2,3,4]\bigg|_{dilaton} &= \frac{(V^{\mu\nu}(p_1,p_2)\cl{E}^{d}_{\mu\nu}(q))(V^{\rho\sigma}(p_3,p_4)\cl{E}^{d}_{\rho\sigma}(q))}{s}\\
&= \frac{i \kappa_D ^2 \left((D-2) s+4 m_1^2\right) \left((D-2) s+4 m_2^2\right)}{16 (D-1) (s-M^2)}, 
\end{align}
which gives a non-relativistic amplitude
\begin{equation}
\cl{M}_4[1,2,3,4]\bigg|_{dilaton,~ M\neq 0}^{NR} = \frac{i \kappa_D ^2 m_1^2 m_2^2}{(D-1) (s-M^2)}.
\end{equation}
\bibliographystyle{JHEP}
\bibliography{3Dbib}
\end{document}